\documentclass[reprint,twocolumn,superscriptaddress,amsmath,amssymb,prl]{revtex4-1}
\usepackage{dcolumn}
\usepackage{epsfig}
\usepackage{graphics}
\usepackage[latin1]{inputenc} 
\usepackage[T1]{fontenc}
\usepackage{bm}
\newcommand{\ddst}{false}

\begin{document}

\title{Topological Origin of Fracture Toughening in Complex Solids: the Viewpoint of Rigidity Theory}

\author{M. Bauchy}
 \email[Contact: ]{bauchy@ucla.edu}
 \homepage[\\Homepage: ]{http://mathieu.bauchy.com}
 \affiliation{Department of Civil and Environmental Engineering, University of California, Los Angeles, CA 90095, United States}
\author{M. J. Abdolhosseini Qomi}
 \affiliation{Concrete Sustainability Hub, Department of Civil and Environmental Engineering, Massachusetts Institute of Technology, 77 Massachusetts Avenue, Cambridge, MA 02139, United State}
\author{C. Bichara}
 \affiliation{Centre Interdisciplinaire des Nanosciences de Marseille, CNRS and Aix-Marseille University, Campus de Luminy, Marseille, 13288 Cedex 09, France}
\author{F.-J. Ulm}
 \affiliation{Concrete Sustainability Hub, Department of Civil and Environmental Engineering, Massachusetts Institute of Technology, 77 Massachusetts Avenue, Cambridge, MA 02139, United State}
\author{R. J.-M. Pellenq}
 \affiliation{Concrete Sustainability Hub, Department of Civil and Environmental Engineering, Massachusetts Institute of Technology, 77 Massachusetts Avenue, Cambridge, MA 02139, United State}
 \affiliation{MIT-CNRS joint laboratory at Massachusetts Institute of Technology, 77 Massachusetts Avenue, Cambridge, MA 02139, United States}
 \affiliation{Centre Interdisciplinaire des Nanosciences de Marseille, CNRS and Aix-Marseille University, Campus de Luminy, Marseille, 13288 Cedex 09, France}

\begin{abstract}

In order to design tougher materials, it is crucial to understand the relationship between their composition and their resistance to fracture. To this end, we investigate the fracture toughness of usual sodium silicate glasses (NS) and complex calcium--silicate--hydrates (CSH), the binding phase of cement. Their atomistic structure is described in the framework of the topological constraints theory, or rigidity theory. We report an analogous rigidity transition, driven by pressure in NS and by composition in CSH. Relying both on simulated and available experimental results, we show that optimally constrained isostatic systems show improved fracture toughness. The flexible to stressed--rigid transition is shown to be correlated to a ductile-to-brittle transition, with a local minimum of the brittleness for isostatic system. This fracture toughening arises from a reversible molecular network, allowing optimal stress relaxation and crack blunting behaviors. This opens the way to the discovery of high-performance materials, designed at the molecular scale.
\end{abstract}


\maketitle

%%%%  INTRODUCTION

Resistance to fracture, or fracture toughness, is of primary importance in material science, with direct application to building construction and manufacturing, e.g., the protective screens for smartphones. Indeed, tougher materials allow increasing mechanical resistance or allow the use of less material, thus achieving the same performances while reducing volumes and weight, which is critical for items such as handheld devices. To design tougher materials, one needs to understand the relationship between toughness and composition. The latter is usually not trivial. In bulk materials, although elastic moduli and hardness typically increase with the average coordination number of the atoms \cite{varshneya_microhardness_2007}, toughness shows non-monotonic behaviors \cite{guin_hardness_2002, varshneya_microhardness_2007}.

In this Letter, we investigate the relationship between fracture toughness and atomic-scale topology by means of molecular dynamics (MD) simulations and topological constraints, or rigidity theory \cite{phillips_topology_1979, phillips_topology_1981, thorpe_continuous_1983, mauro_topological_2011, bauchy_topological_2012}. The latter, inspired by Maxwell's study on the stability of mechanical trusses \cite{lagrange_mecanique_1864}, has been extensively used to understand the compositional dependence of network glasses while only relying on an atomic-scale analysis. A molecular network can be classified as \textit{flexible}, having internal degrees of freedom called floppy modes \cite{naumis_energy_2005} that allow for local deformations; \textit{stressed--rigid}, being locked by its high connectivity; or \textit{isostatic}, the optimal intermediate state. The isostatic state is achieved when the number of constraints per atom $n_c$, comprising radial bond-stretching (BS) and angular bond-bending (BB), equals 3, the number of degrees of freedom per atom. Compositions offering an isostatic behavior have been found to exist inside a window \cite{feng_direct_1997}, located between the flexible ($n_c<3$) and the stressed--rigid ($n_c>3$) compositions, known as the Boolchand intermediate phase (IP), and show some remarkable properties such as a stress-free character \cite{chubynsky_self-organization_2006}, a space-filling tendency \cite{rompicharla_abrupt_2008}, weak aging phenomena \cite{chakravarty_ageing_2005}, and anomalous dynamical and structural signatures \cite{bauchy_transport_2013, bauchy_compositional_2013, micoulaut_anomalies_2013, bauchy_percolative_2013}. Rigidity theory has been shown to be a powerful tool for predicting the mechanical properties of glass \cite{smedskjaer_prediction_2010, smedskjaer_impact_2010} and has been used to design the Gorilla$^{\circledR}$ Glass 3 \cite{mauro_glass:_2013, mauro_unified_2012} from Corning$^{\circledR}$, a high-performance glass used for most smartphones and tablets screens.

We recently showed that rigidity theory could be extended to handle calcium--silicate--hydrates (CSH) \cite{bauchy_extension_2014}. CSH is the binding phase of cement, a ubiquitous material in our built environment, and is believed to be responsible for its mechanical properties \cite{allen_composition_2007}. It can be classified as complex in the sense that it is anisotropic, partially crystalline and heterogeneous, and contains some free water molecules \cite{pellenq_realistic_2009, manzano_confined_2012, abdolhosseini_qomi_concrete_2013, qomi_anomalous_2014, bauchy_order_2014}. On the other hand, sodium silicate (NS) is a base material of various multi-component silicate glasses found in industry (e.g., window glasses) as well as in geoscience (e.g., magmas \cite{bauchy_viscosity_2013}). Here, we present a combined analysis of the fracture toughness of those materials, based on molecular dynamics simulations, in the framework of rigidity theory.

%%%% SIMULATION DETAILS

To investigate fracture properties and the topology of the atomic network, one needs realistic MD simulations, i.e., realistic potentials. The potential used for the NS glasses we have investigated has been extensively studied and has been shown to provide realistic results for structure, dynamics, and mechanics \cite{du_medium_2004, bauchy_pockets_2011, bauchy_structural_2012, bauchy_viscosity_2013, pedone_insight_2007}. We have simulated 2SiO$_2$--Na$_2$O glasses made of 3000 atoms, which have been permanently densified via a cooling at constant selected pressures. At ambient pressure, pure silica consists of a network of 4-fold coordinated silicon tetrahedra interconnected by 2-fold bridging oxygen atoms (BO). In NS, sodium atoms tend to depolymerize the network by breaking Si--BO--Si bonds, thus creating non-bridging oxygen atoms (NBO). On the contrary, the pressure experienced during the cooling tends to increase the coordination number of Si and O atoms, thus increasing the rigidity of the network. More details about the simulation procedures and the influence of pressure on the structure have been reported previously \cite{bauchy_atomic_2011, bauchy_pockets_2011, bauchy_structural_2012, bauchy_topological_2012, bauchy_viscosity_2013, bauchy_transport_2013, micoulaut_anomalies_2013, bauchy_percolative_2013}. Prior to fracture simulations, the obtained NS glasses are relaxed to zero pressure, but they remain permanently densified, with highly coordinated network formers atoms.

Surprisingly, the structure of CSH at the atomic-scale has come to light only recently. To describe its disordered molecular structure, Pellenq et al. \cite{pellenq_realistic_2009} proposed a realistic model for CSH with the stoichiometry of (CaO)$_{1.65}$(SiO$_2$)(H$_2$O)$_{1.73}$. Following the same methodology, we have generated CSH models of different compositions by introducing defects in an 11 \AA\ tobermorite \cite{hamid_crystal-structure_1981} configuration, following a combinatorial procedure. 11 \AA\ tobermorite consists of pseudo-octahedral calcium oxide sheets, which are surrounded by silicate tetrahedral chains. The latter are made of BO atoms and $Q^2$ silicon atoms \cite{abdolhosseini_qomi_evidence_2012}. Those negatively charged calcium--silicate sheets are separated from each other by an interlayer spacing, which contains water molecules and charge-balancing calcium cations. Whereas the Ca/Si ratio in 11 \AA\ tobermorite is 1, this ratio is increased up to 1.9 in the present CSH models through randomly removing SiO$_2$ groups. The defects in silicate chains provide possible sites for the adsorption of extra water molecules. The adsorption of water molecules in the structurally defected tobermorite model was performed via the Grand Canonical Monte Carlo method, ensuring equilibrium with bulk water at constant volume and room temperature. The REAXFF potential \cite{manzano_confined_2012, qomi_anomalous_2014}, a reactive potential, was then used to account for the reaction of the interlayer water with the defective calcium--silicate sheets. The use of the reactive potential allows observing the dissociation of water molecules into hydroxyl groups. More details on the preparation of the model and on the multiple validations with respect to experiments can be found in previous publications \cite{pellenq_realistic_2009,abdolhosseini_qomi_concrete_2013,qomi_anomalous_2014}.

%%%% CONSTRAINTS ENUMERATION METHOD

\begin{figure}
\includegraphics*[height=\linewidth, angle=-90, keepaspectratio=true, draft=\ddst]{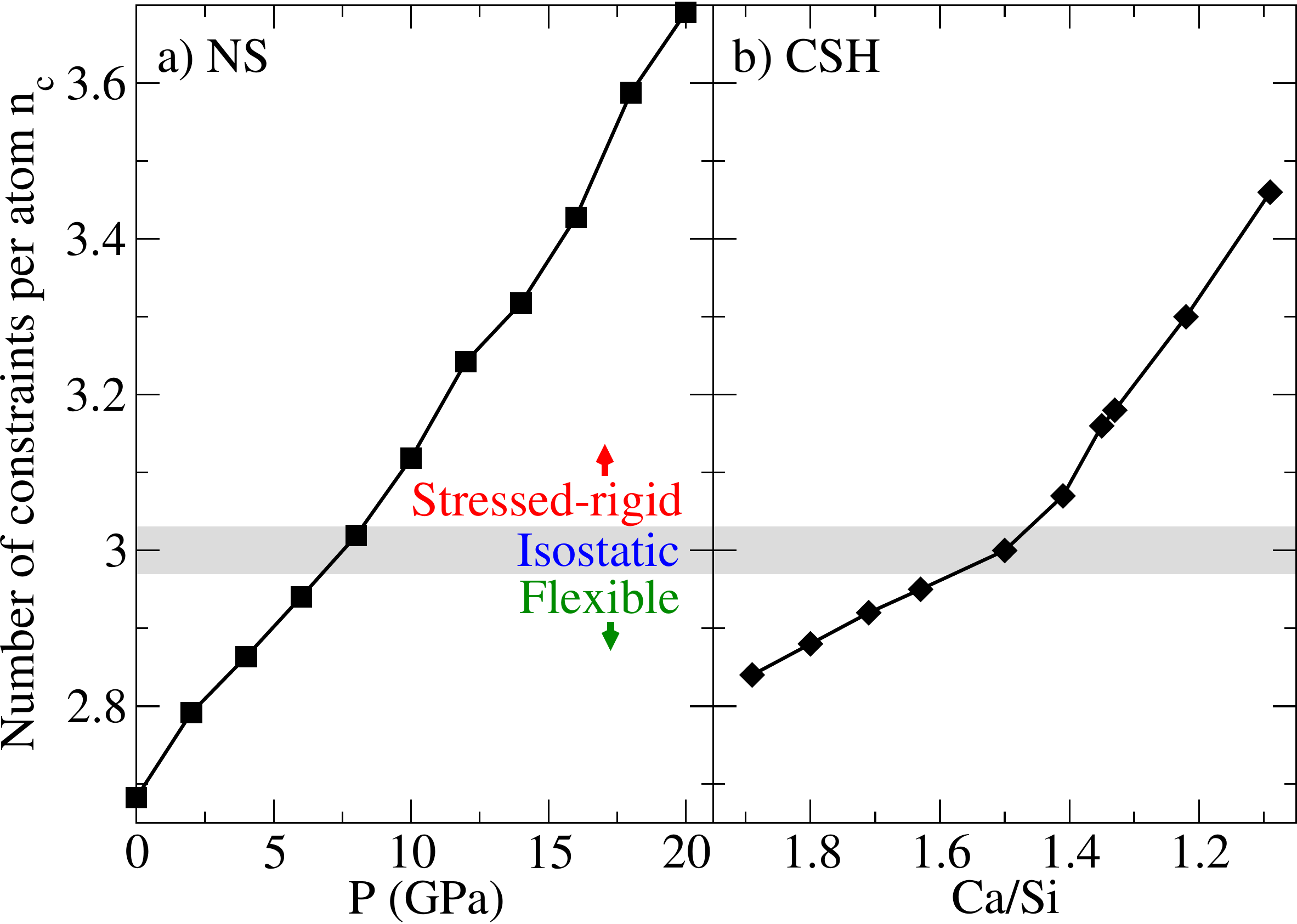}
\caption{\label{fig:nc} (Color online) Number of constraints per atom in (a) NS and (b) CSH as a function of pressure and Ca/Si molar ratio, respectively. The gray area is an approximate boundary between the flexible and the stressed--rigid regimes ($n_c$=3).
}
\end{figure}

The enumeration of the mechanical constraints experienced by the atoms inside a molecular network requires care. Although the counting is obvious in fully connected glass like Ge$_x$Se$_{1-x}$ (GS), in which $n_c=5x+2$ \cite{bauchy_angular_2011}, enumerating topological constraints in more complex materials is not as straightforward, as their coordination numbers are not necesserily known. Moreover, it has been shown that the constraints can be intact or broken according to the temperature \cite{smedskjaer_quantitative_2010, bauchy_atomic_2011}. To tackle this issue,  we refined a method originally developed in Ref. \cite{micoulaut_understanding_2010}, and widely applied to chalcogenide and oxide glasses since then \cite{bauchy_angular_2011, bauchy_atomic_2011, bauchy_topological_2012, micoulaut_anomalies_2013, bauchy_compositional_2013, bauchy_transport_2013, bauchy_structure_2013, bauchy_structure_2013-1, bauchy_percolative_2013, micoulaut_structure_2013}. The latter is based on the analysis of atomic trajectories obtained through molecular dynamics simulations, from which the numbers of BS and BB constraints are computed by analyzing the standard deviations $\sigma$ of the bond lengths and angular distributions, respectively. The basic idea is intact constraints induce low $\sigma$, whereas broken ones result in large $\sigma$. The details of the constraints enumeration in NS and CSH can be found in Ref. \cite{bauchy_transport_2013, bauchy_extension_2014}.

%%%% CONSTRAINTS ENUMERATION RESULTS

As shown in Fig. \ref{fig:nc}, both NS and CSH show a rigidity transition. As previously reported \cite{bauchy_transport_2013}, NS becomes stressed--rigid at high pressure, due to the increase of the coordination number of Si and O atoms (see Fig. \ref{fig:nc}a). Similarly to GS, CSH shows a rigidity transition according to composition, being flexible ($n_c<3$) at high Ca/Si ratio, stressed--rigid ($n_c>3$) at low Ca/Si ratio and isostatic ($n_c=3$) at Ca/Si$\simeq$1.5. This result constitutes the first direct evidence of composition-driven rigidity transition in a complex material. It is worth noticing that the Ca/Si$\simeq$1.5 transition composition also corresponds to structural and mechanical transitions, as the system is crystalline and transversely isotropic at low Ca/Si ratios and amorphous and isotropic at high Ca/Si ratios \cite{abdolhosseini_qomi_concrete_2013}. This feature is very similar to what is typically observed in glasses, as stressed--rigid compositions show a weak glass-forming ability since they tend to easily crystalize \cite{mauro_topological_2011}. The trend of $n_c$ shows a bilinear behavior with Ca/Si, the slope being shallower in the flexible domain. This can be interpreted as flexible systems demonstrating an improved ability to self-organize and go toward the isostatic optimal state, as opposed to the locked crystalline systems at Ca/Si$<$1.5.

%%%% FRACTURE METHODOLOGY

\begin{figure*}
\includegraphics*[width=0.9\linewidth, keepaspectratio=true, draft=\ddst]{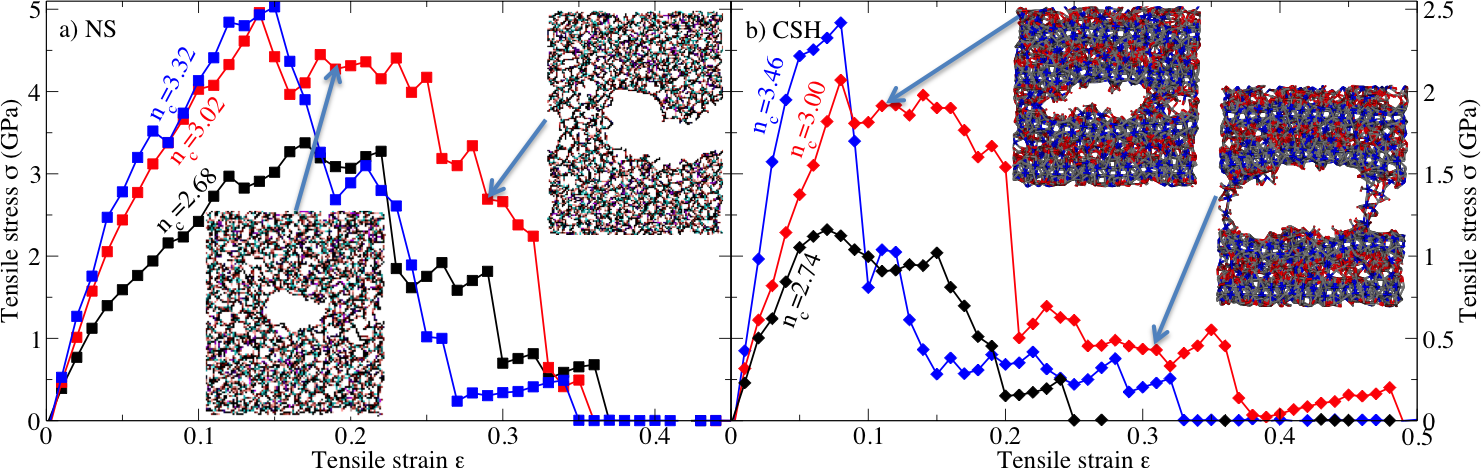}
\caption{\label{fig:ss} (Color online) Tensile stress as a function of the tensile strain in (a) NS and (b) CSH for three systems: flexible, isostatic, and stressed--rigid. Snapshots of the molecular configurations at different strains are displayed.
}
\end{figure*}

We now simulate the fracture behavior in mode I of the NS and CSH systems. To this end, we have followed the method introduced by Brochard et al. \cite{brochard_fracture_2013}, which studies the fracture properties at the smallest scales based on molecular dynamics simulations. This approach relies on the energetic theory of fracture mechanics \cite{griffith_phenomena_1921, leblond_mecanique_2003, anderson_fracture_2005}. We first insert a sharp initial crack, expected to appear naturally, e.g., during the cooling process. After a relaxation to zero pressure, the system is elongated stepwise in the weakest direction  $z$ by small increments of the strain $\epsilon$. At each step, after an initial equilibration, the stress along the $z$ axis $\sigma_z$ is averaged. Once the system is broken, the stress goes back to zero, so that the fracture energy $ G_c$ can be estimated by integrating the stress over the strain. Alternatively to the energetic approach, the notion of fracture toughness $K_{Ic}$ is usually used in engineering applications. This quantity was introduced by Irwin \cite{irwin_fracture_1958} as the maximum stress intensity at the crack tip that a solid can undergo, and below which propagation cannot occur. The relationship between $K_{Ic}$ and $G_c$ is given by the Irwin formula \cite{irwin_fracture_1958}. The details of the methodology can be found in Ref. \cite{bauchy_fracture_2014}.

%%%% FRACTURE RESULTS

Fig. \ref{fig:ss} shows $\sigma_z$ with respect to the tensile strain $\epsilon$ both for NS and CSH, in the case of three flexible, isostatic and stressed--rigid systems. At low strain, the mechanical response is linear elastic. The stress thus increases linearly with the strain up to a maximum value, the slope being related the Young's modulus of the system. During this stage, the crack does not propagate and the free energy of the system is stored in the form of mechanical elastic energy only. At larger strain, the crack starts to propagate. As opposed to brittle materials like quartz \cite{brochard_fracture_2013}, both systems show some ductility in the sense that the crack does not propagate instantly after a given critical strain. Thanks to its internal flexibility, the network rather deforms to prevent the fracture from occurring, as can be observed in the snapshots inside Fig. \ref{fig:ss}. 

\begin{figure}
\includegraphics*[width=\linewidth, keepaspectratio=true, draft=\ddst]{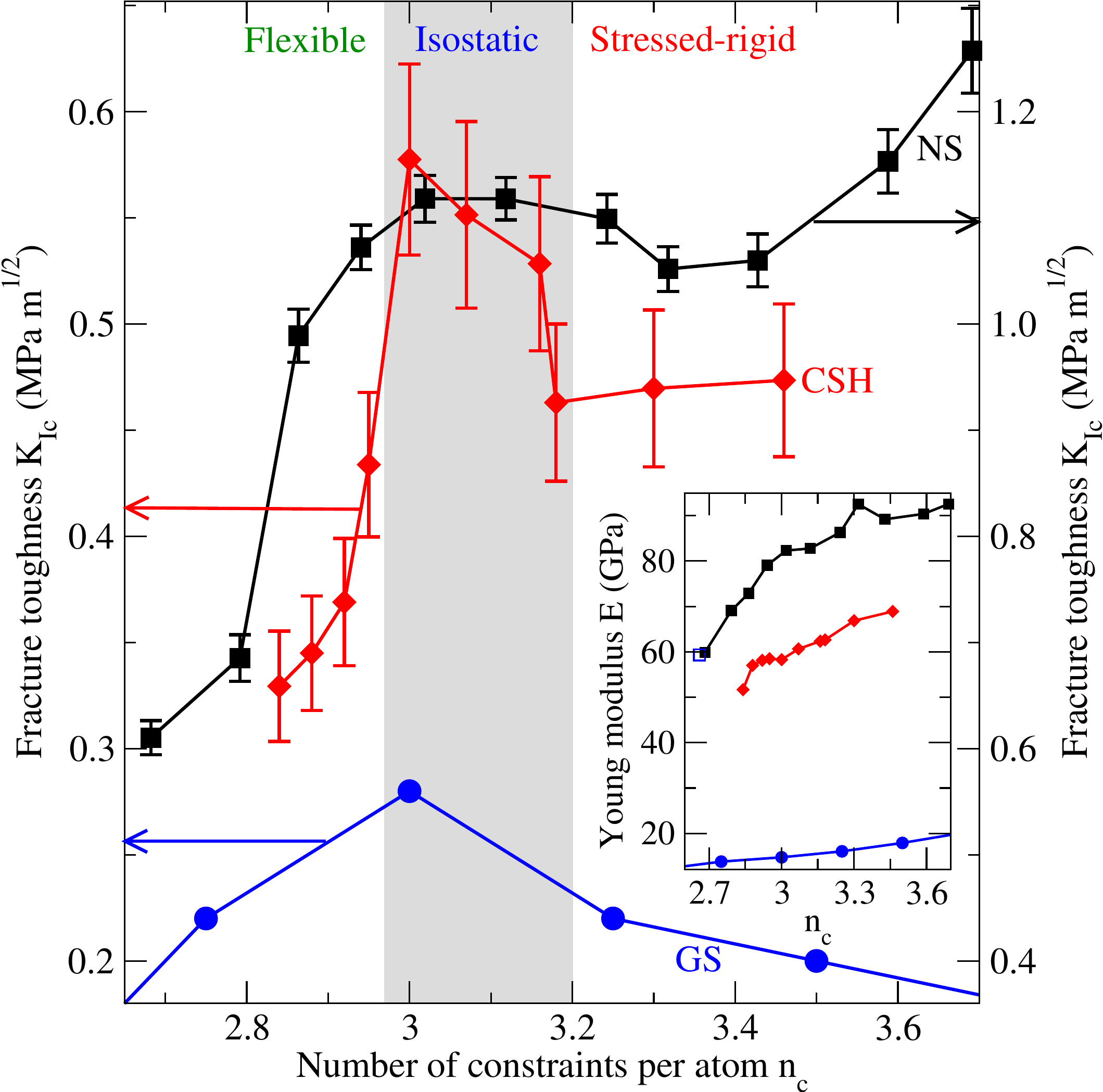}
\caption{\label{fig:Kc} (Color online) Fracture toughness $K_{Ic}$ as a function of the number of constraints per atom $n_c$ for NS, CSH (computed), and GS (experiment \cite{guin_hardness_2002}). The gray area highlights an intermediate window where $K_{Ic}$ is maximal. The inset shows the Young's modulus as a function of $n_c$ for NS, CSH (computed, compared with an experimental value at ambient pressure for NS, empty square \cite{bansal_handbook_1986}), and GS (experimental \cite{guin_hardness_2002}).
}
\end{figure}

Fig. \ref{fig:Kc} shows $K_{Ic}$, obtained from an integration of the previous stress--strain curves and according to $n_c$, for NS and CSH. Experimental values for GS \cite{guin_hardness_2002} are added for comparison. For NS at ambient pressure, we find $K_{Ic}$=0.61 MPa m$^{1/2}$ and $G_c$=5.9 J/m$^2$, which is in fair agreement with the experimental value of 7$\pm$1 J/m$^2$ \cite{kennedy_strength_1973}. For CSH at Ca/Si = 1.71, we find $G_c = 1.7$ J/m$^2$ and $K_{Ic} = 0.37$ MPa.m$^{1/2}$, which is in the typical range of values of cement paste toughness \cite{bauchy_fracture_2014}. Interestingly, for all the three systems, we observe a maximum of the resistance to fracture inside a window, approximately located between $n_c$=3 and 3.2, close to the isostatic threshold $n_c$=3. This anomaly cannot be explained by a maximum of the elastic modulus (see the inset of Fig. \ref{fig:Kc}). Rather, as it can be seen on the stress--strain curves (see Fig. \ref{fig:ss}), this anomaly arises from an increased ability to reach high strain before the crack propagation. To our knowledge, this is the first time that simulation succeeds to capture the fracture toughening of isostatic networks observed experimentally for GS \cite{guin_hardness_2002, varshneya_microhardness_2007}. 

%%%% FRACTURE DISCUSSION

\begin{figure}
\includegraphics*[height=\linewidth, angle=-90, keepaspectratio=true, draft=\ddst]{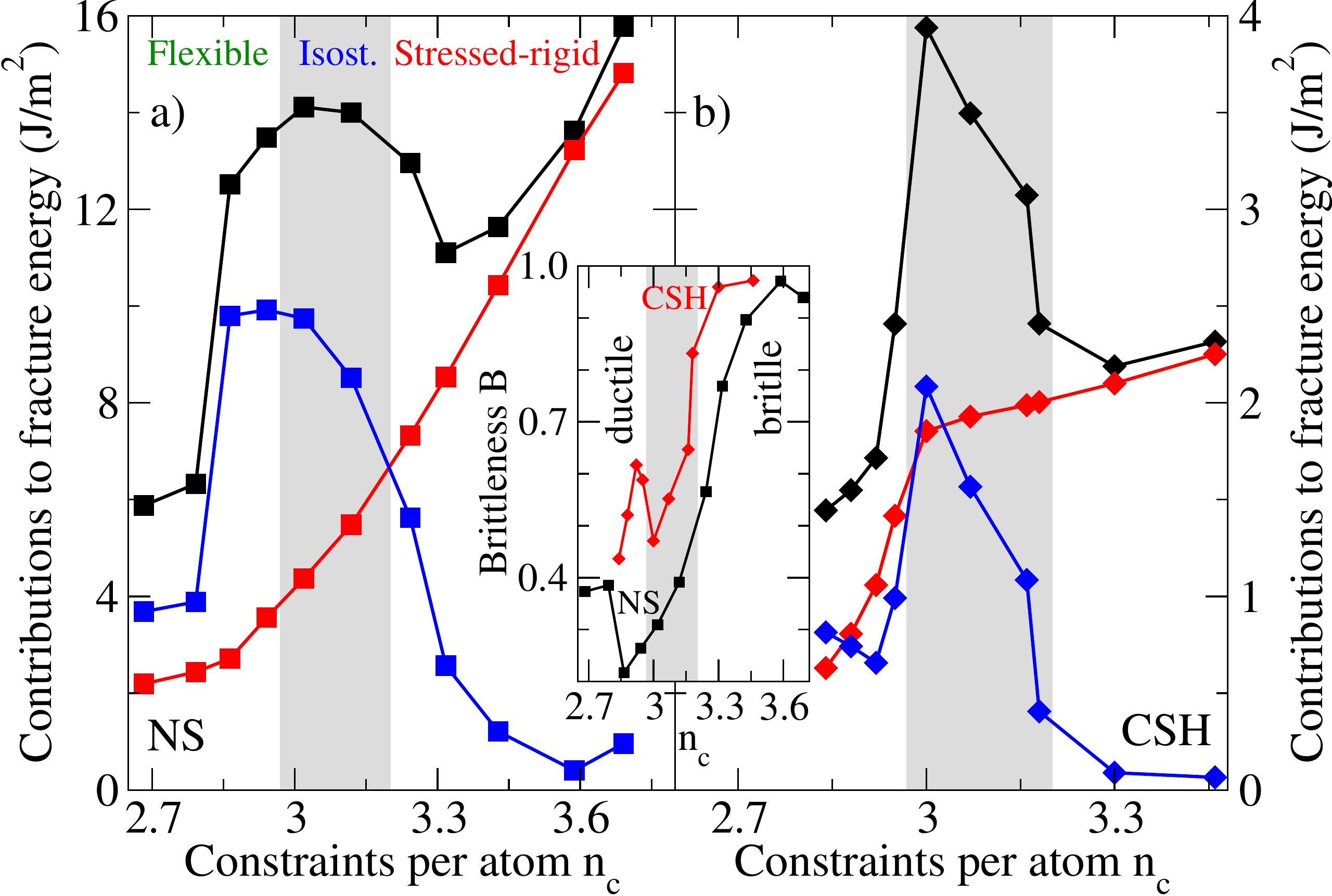}
\caption{\label{fig:Gc} (Color online) Fracture energy (black), surface energy (doubled, red), and dissipated energy (blue) for (a) NS and (b) CSH as a function of the number of constraints per atom $n_c$. The inset shows the brittleness parameter as a function of $n_c$. The gray areas highlight the intermediate window where the fracture toughness is maximal.
}
\end{figure}

We now investigate the origin of this observed toughening by evaluating the different contributions to the fracture energy. $G_c$ is linked to the surface energy $\gamma_s$ by $G_c = 2\gamma_s + G_{\text{diss}}$, where $G_{\text{diss}}$ captures all forms of dissipated energy and is equal to zero for a perfectly brittle material. Here, $\gamma_s$ was roughly estimated from MD simulation by cutting the system into two parts among the plane of the fracture, letting it relax and computing the change of its potential energy (see more details in Ref. \cite{bauchy_fracture_2014}). Fig. \ref{fig:Gc} shows $G_c$, $2\gamma_s$, and $G_{\text{diss}}$. First, we note that, both for NS and CSH, $\gamma_s$ monotically increases with $n_c$. This is not surprising, as $\gamma_s$ is the energy needed to break the chemical bonds and should, therefore, increase with the number of constraints per unit of surface. Secondly, we realize that the fracture of NS at ambient pressure shows some ductility. This is in agreement with recent experiments \cite{celarie_glass_2003, livne_breakdown_2008} and simulations \cite{chen_interaction_2007} suggesting that, although glasses are brittle at the macroscale, they show some nanoductility. Hence, the toughening of isostatic systems arises from some kind of energy dissipation. The brittleness $B$ can be evaluated by $B=2\gamma_s/G_c$, which is equal to 1 for a perfectly brittle material. For both NS and CSH, we note that the flexible/stressed--rigid transition is fairly well correlated to a ductile/brittle transition, although the isostatic threshold shows a local minimum of brittleness. This feature is similar to what is observed experimentally in GS \cite{guin_hardness_2002, varshneya_microhardness_2007}, so that it could be a general feature of the rigidity transition.

%%%% ENERGY DISSIPATION

\begin{figure}
\includegraphics*[height=\linewidth, angle=-90, keepaspectratio=true, draft=\ddst]{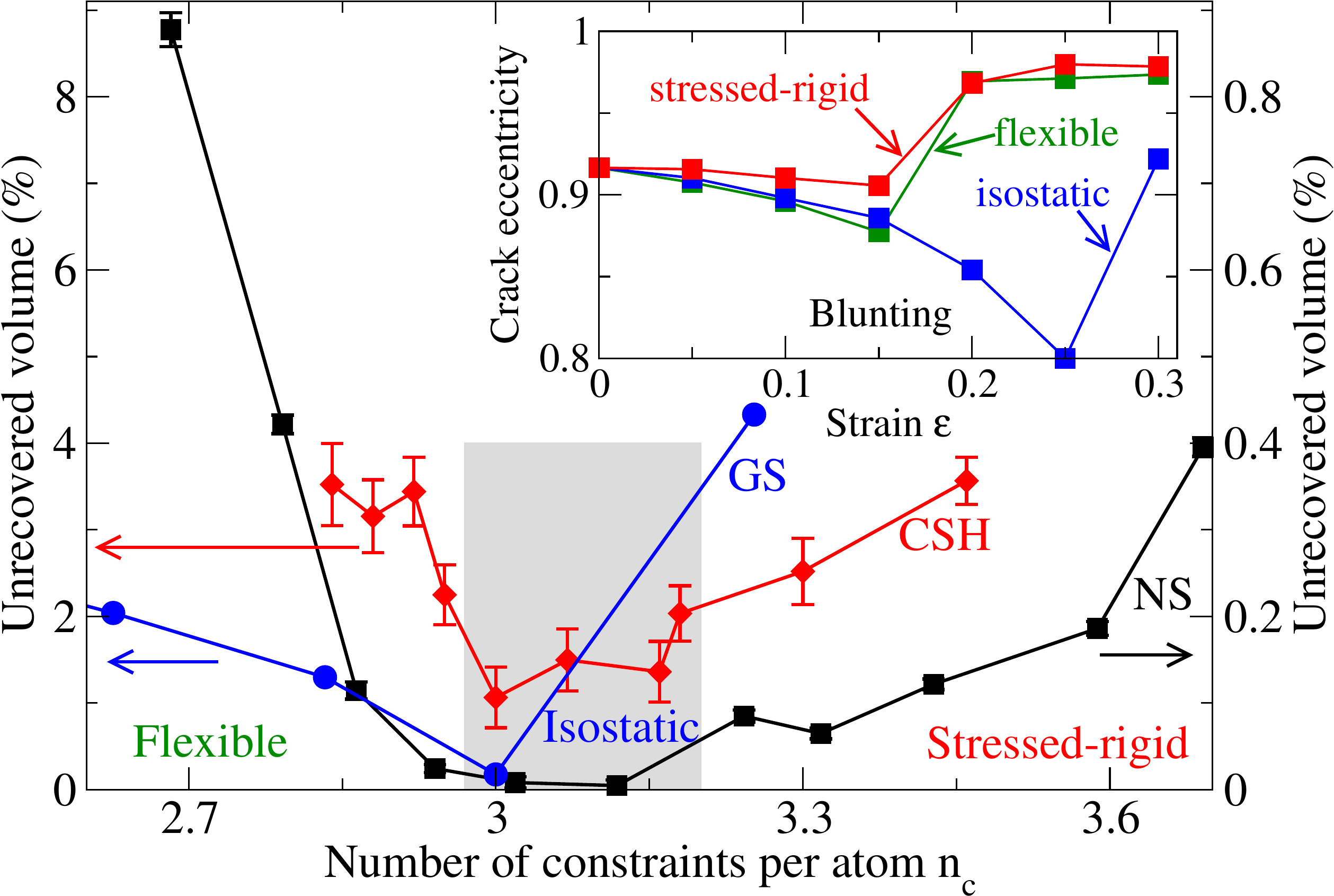}
\caption{\label{fig:rever} (Color online) Fraction of unrecovered volume after loading and unloading for NS, CSH, and GS (from Ref. \cite{mauro_modeling_2007}). The gray area highlights the intermediate window where the fracture toughness is maximal. The inset shows the eccentricity of the elliptic crack as a function of strain for three NS systems: flexible, isostatic, and stressed--rigid.
}
\end{figure}

The dissipation of energy during fracture can take different forms, such as non-reversible plasticity, crack blunting due to visco-elastic stress relaxation, or zig--zag crack paths. We first investigate the plasticity of the network. To this end, we have loaded NS and CSH systems at 10 GPa during 1 ns before unloading them and computing the relative unrecovered volume. Interestingly, as shown in Fig. \ref{fig:rever}, we observe that, inside the previously defined window, the network is reversible with pressure changes, as the volume is close to be fully recovered. In that sense, this window shares a strong analogy with the Boolchand IP, in which systems show a reversible glass transition with temperature. The reversibility with pressure of isostatic networks has already been observed for GS \cite{varshneya_microhardness_2007, mauro_modeling_2007} and is explained as follows. Thanks to their internal floppy modes, flexible systems can easily undergo irreversible deformations during the loading. On the contrary, stressed--rigid systems are completely locked and remain frustrated in their pressurized configuration. Eventually, isostatic systems simply adapt with pressure in a reversible way. Once again, this reversibility with pressure could be a generic signature of isostatic systems.

%%%% FINAL DISCUSSION

Hence, plasticity does not explain the observed toughening, as it is, on the contrary, minimal for isostatic systems. It appears that it is the reversibility with stress of the network that allows it to resist fracturing. Indeed, for isostatic systems, we observe a crack blunting behavior, the latter showing large deformation of its shape before any propagation (see the snapshots in Fig. \ref{fig:ss}). With this miss of sharpness of the crack, the stress is not concentrated any more at the crack tips, which requires a higher stress to allow the propagation. This manifests as a plateau in the stress--strain curves (see Fig. \ref{fig:ss}). To quantify the extent of the blunting, we fitted to the crack surface, at different strain, an elliptic cylinder and calculated its eccentricity (0 for a circle, 1 for a parabola, see the inset of Fig. \ref{fig:rever}). First, we note that the crack blunts by itself at $\epsilon=0$, as an infinitely sharp crack is not stable at the atomic-scale. As $\epsilon$ increases, the crack slightly blunts as the system elastically deforms. For flexible and stressed--rigid systems, the crack becomes sharper as soon as it starts to propagate. On the contrary, for the isostatic system, the crack continues to blunt, up to $\epsilon=0.25$. This arises from an optimal relaxation of stress by the network, which is supported by the fact that the isostatic NS systems show a minimum of viscosity and a minimum of activation energy \cite{bauchy_transport_2013}. The collective shearing motion involved during the crack blunting requires both flexibility, to allow the deformation, and some rigidity, to allow a rearrangement over large scales. Hence, this toughening is only possible in isostatic networks, which are rigid, but stress-free.

%%%% CONCLUSION

It is striking to see that, although belonging to different families of materials, NS, CSH, and GS show a similar relation between their atomic topology and their resistance to fracture. This result suggests that all isostatic systems, characterized by a rigid but stress-free network, are likely to show reversibility with pressure and toughening. This study highlights that the design of isostatic materials from the molecular scale could lead to a greater resistance to fracture, thus allowing the use of less material. The toughening may also be linked to the weak aging phenomena that isostatic systems are reported to undergo \cite{chen_long_2010}. This will be the subject of further investigations. More generally, we are convinced that applying glass science tools to the wider area of complex materials will lead to a better understanding of their nature and of the influence of their nanoscale composition.

\begin{acknowledgments}
This work has been carried out within the framework of the ICoME2 Labex (ANR-11-LABX-0053) and the A*MIDEX projects (ANR-11-IDEX-0001-02) cofunded by the French program "Investissements d'Avenir" which is managed by the ANR, the French National Research Agency. It was also supported by Schlumberger under an MIT-Schlumberger research collaboration and by the CSHub at MIT. 
\end{acknowledgments}

%\bibliography{bauchy}

\end{document}